\documentclass[letterpaper]{article}
\usepackage{aaai}
\usepackage{times}
\usepackage{helvet}
\usepackage{courier}
\usepackage{textcomp}
\usepackage{xcolor}

\usepackage{booktabs} 

\usepackage{mathrsfs}

\usepackage{amsmath,amsfonts}
\usepackage{bm}

\usepackage{graphicx} 
\usepackage{subfigure}

\usepackage[english]{babel}
\usepackage{amsmath}
\usepackage[colorinlistoftodos]{todonotes}
\usepackage{algorithm}
\usepackage{algorithmic}
\usepackage{graphicx}
\usepackage{multirow}
\usepackage{booktabs}
\usepackage{comment}

\AtBeginDocument{%
  \providecommand\BibTeX{{%
    \normalfont B\kern-0.5em{\scshape i\kern-0.25em b}\kern-0.8em\TeX}}}

\begin{document}

\title{Attributed Graph Neural Networks for Recommendation Systems \\on Large-Scale and Sparse Graph
}


\author{Yifei Zhao, Mingdong Ou, Rongzhi Zhang, Meng Li\\ Alibaba Group, China \\ \{andy.zyf, mingdong.omd, rose.zrz, molly.lm\}@alibaba-inc.com}
\maketitle

\begin{abstract}
\begin{quote}
Link prediction in structured-data is an important problem for many applications, especially for recommendation systems. Existing methods focus on how to learn the node representation based on graph-based structure. High-dimensional sparse edge features are not fully exploited. Because balancing precision and computation efficiency is significant for recommendation systems in real world, multiple-level feature representation in large-scale sparse graph still lacks effective and efficient solution. In this paper, we propose a practical solution about graph neural networks called \textbf{A}ttributed \textbf{G}raph \textbf{C}onvolutional \textbf{N}etworks(AGCN) to incorporate edge attributes when apply graph neural networks in large-scale sparse networks. We formulate the link prediction problem as a subgraph classification problem. We firstly propose an efficient two-level projection to decompose topological structures to node-edge pairs and project them into the same interaction feature space. Then we apply multi-layer GCN to combine the projected node-edge pairs to capture the topological structures. Finally, the pooling representation of two units is treated as the input of classifier to predict the probability. We conduct offline experiments on two industrial datasets and one public dataset and demonstrate that AGCN outperforms other excellent baselines. Moreover, we also deploy AGCN method to important scenarios on Xianyu and AliExpress. In online systems, AGCN achieves over 5\% improvement on online metrics.
\end{quote}
\end{abstract}

\begin{figure}
    \centering
    \includegraphics[width=0.9\linewidth]{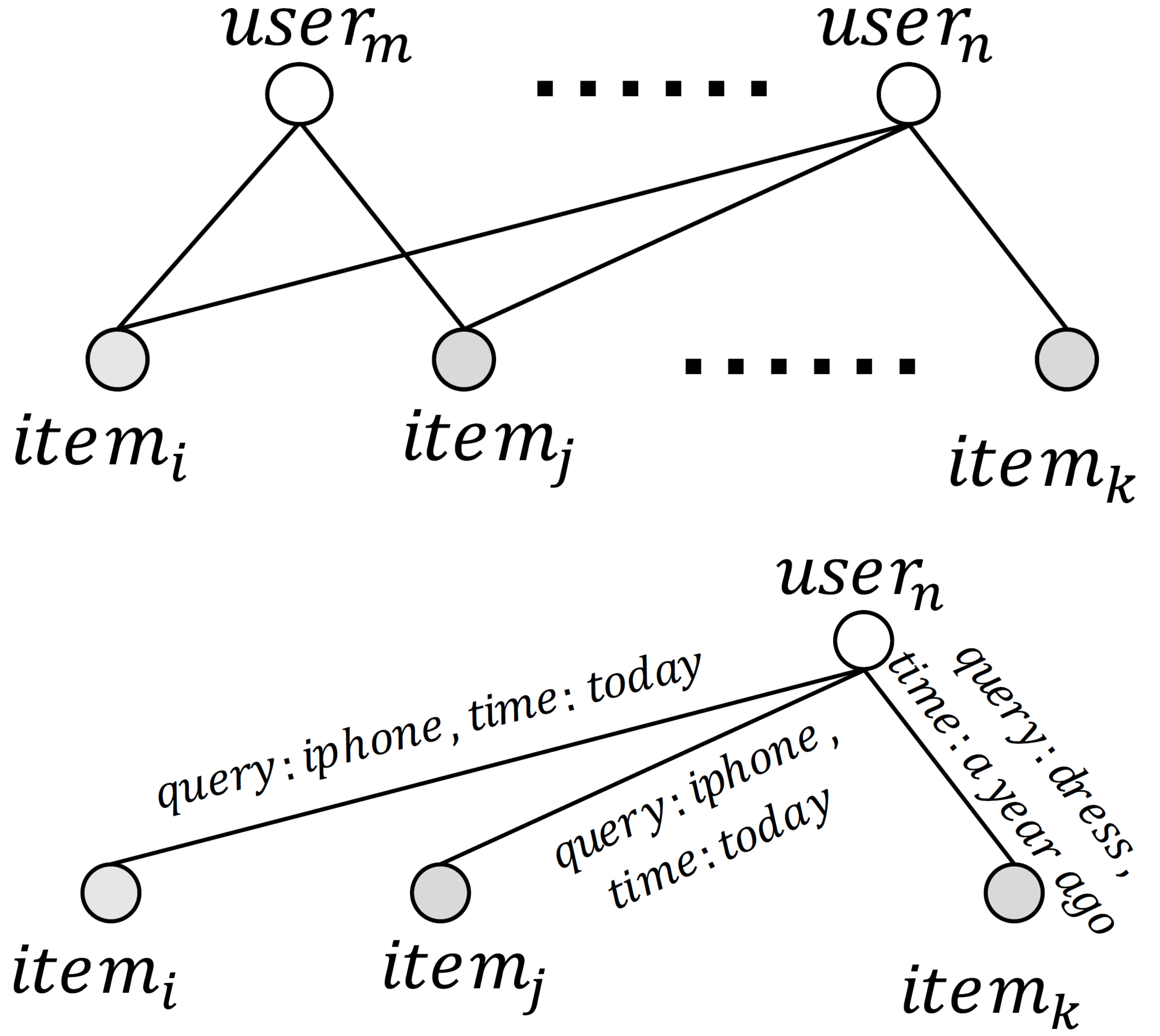}
    \caption{The influence of edge attributes on user-item bipartite graph.}
    \label{fig:motivation}
\end{figure}

\section{Introduction}
Link prediction problem seeks to predict the probability of the link between two target nodes. Many applications can be formulated as link prediction problems. For example, personalized item recommendation can be formulated as link prediction between users and items. User-item interactions are constructed as a bipartite graph shown in Figure 1. Similar item recommendation can be treated as predicting the link between two items and friend recommendation can be treated as predicting the link between two users.

Due to the wide application of this task, link prediction problem in graph-structured data has attracted much attention in research and industry. 
Traditional methods in industry are heuristic methods, because they are interpretable and easy to implement \cite{zhang2018link}. They use handcrafted measures of topological structure to estimate the link probability between nodes. Most methods assume that more paths between nodes lead to the higher the link probability. Heuristic methods suffer bad performances when the assumptions are not satisfied. Moreover, when the graph is very sparse, most of target nodes are connected with few paths. sparse links cannot be distinguished. Instead of handcrafted measures, supervised methods learn the topological features through complex models.  Weisfeiler-Lehman Neural Machine (WLNM) is the first supervised method\cite{zhang2017weisfeiler}. SEAL~\cite{zhang2018link} introduce GCN to extract topological features for link prediction. 

However, all the methods mentioned above ignore abundant edge attributes, which is very important to link prediction, especially in sparse graphs. Taking Figure 1. as an example, it is extracted from Figure 1. with edge attributes. User$_{n}$ searches iphone today and clicks item$_{i}$ and item$_{j}$ and a year ago User$_{n}$ searches dress and clicks item$_{k}$. Ignoring the edge attributes, item$_{i}$-item$_{j}$ and item$_{i}$-item$_{k}$ are both connected by one path, and their link cannot be distinguished. Considering the edge attributes, as user$_{n}$-item$_{i}$ and user$_{n}$-item$_{j}$ share the same query and interaction time, item$_{i}$ is intuitively more similar with item$_{j}$ than item$_{k}$. Meanwhile, sparse graph is common in practice, and edge attributes are not fully exploited in extracting the topological structures. Therefore, edge attributes have the potential to improve the accuracy of link prediction in practical applications.

In this paper, we propose a practical solution based on GCN, called {\it Attributed Graph Convolution Network} (AGCN), to introduce both edge and node attributes for link prediction. We first treat each node-edge pair as a base element and project corresponding edge and node attributes into the same feature space. As the attribute vectors are often very high-dimensional in practice, the number of parameters of the projection can be very large. That will make the cost of training prohibitive. We formulate the link prediction problem as a variation of subgraph classification problem, where the subgraph is the enclosed subgraph around the link to predict. We propose an efficient two-level projection ({\it i.e.} high-level projection and low-level projection) to address the problem. The high-level projection only operates on the high frequency attributes whose number will be much smaller, and we adopt a simple latent factor model to learn the high-level interaction between nodes and edges. The low-level projection operates on all the attributes and adopts non-parametric operations to extract low-level interaction between node and edge. Then we apply multi-layer GCN to combine the projected node-edge pairs to represent the topological structures. Finally, we concatenate the output of two units after pooling layer and use the classifier to predict the link. 

We conduct offline experiments on public datasets and million-scale datasets derived from real industry. The results confirm that edge attributes help improve the accuracy of link prediction. Further more, we deploy the method on the following four recommendation scenarios in industry. AGCN stably outperforms the baseline and demonstrates that AGCN is effective in industrial sparse graphs.

Our contributions can be summarized as follows:
\begin{itemize}
	\item We propose AGCN to apply subgraph-based graph neural network with heterogeneous information for the large-scale and sparse network.
	\item We propose an efficient solution, including the two-level projection and subgraph classification to control the complexity and accelerate the training. We successfully implement the framework in real large-scale applications.
	\item We deploy the method on four very large scale practical applications and achieve significant online improvement.
\end{itemize}

\section{Related Works}\label{sec:relatedwork}
Link prediction is an important problem which attracted many researchers in graph mining area ~\cite{pech2019link,liben2007link,zhang2017weisfeiler,huo2018link,lichtenwalter2010new,song2015efficient,backstrom2011supervised}.We discuss three mainstream method, heuristic method, latent node representation and graph neural networks in the following section.
\subsection{Heuristic Method}
Heuristic methods stem from the hypothesis that if two nodes are close in the network or connected with the same nodes, they are similar. These methods define distance functions to compute the likelihood of links, such as common neighbors, Katz index and Jaccard\cite{rossi2019higher,li2020friend,zhang2018link}. 
Common neighbors assumes more common 1st-order neighbors represent the larger probability of positive relations between two nodes. Multiple hops are also considered by Adamic-Adar, Katz, PageRank and
SimRank~\cite{brin2012reprint,jeh2002simrank}. 
Adamic-Adar is more widely used in industry due to its simplicity. 
However, the distance functions have a limited capacity to capture the complex or sparse interactions in the large scale network.  

\subsection{Latent Node Representation}
latent feature methods factorize interaction matrix to learn latent representation for each node. Matrix factorization (MF)\cite{mnih2008probabilistic} and stochastic block model (SBM)\cite{airoldi2008mixed} are typical methods in link prediction. With the popularity of network embedding methods, node2vector (N2V)\cite{grover2016node2vec}, deepwalk \cite{perozzi2014deepwalk}, LINE \cite{tang2015line}, Struc2Vec ~\cite{ribeiro2017struc2vec} are proposed to implicitly learn low-dimension node representations. They uses random walk to generate contexts of a vertex. Similar contexts means two nodes are closer together in the feature space. Structural context information is also encoded into node representations by ~\cite{ribeiro2017struc2vec}. However, the shallow neural networks have limited capacity to learn node representations. 


\subsection{Graph Neural Networks}
Graph Neural Networks(GNN), which can capture both graph structure and nodes' attributes in the network, have shown its superiority in link prediction task ~\cite{GCN,graphsage,ying2018graph,ng2002spectral}. 
Different from heuristic method, supervised methods learn a model from the labeled data to predict the link score, which get rid of the assumptions.
supervised learning link prediction from the view of graph is firstly proposed by\cite{zhang2017weisfeiler}.  Weisfeiler-Lehman Neural Machine (WLNM) extracts subgraphs from the global network and obtain the subgraph representation through a fully connected neural network. WLNM lacks structure information due to different subgraphs are truncated into the same size. It cannot learn from the explicit and implicit feature representations directly.
To address the above limitation,
Subgraphs, Embeddings and Attributes for
Link prediction(SEAL) is proposed in \cite{zhang2018link}, which makes full use of latent and explicit node attributes and take topological structures into consideration. SEAL achieves the state-of-the-art results. 
IGMC ~\cite{IGMC} applied the SEAL into user-item bipartite graph in recommendation and process edge features like nodes. We noticed that recently most researches introduce edge attributes into graph neural networks ~\cite{talp,gcn-lase,IGMC}, but few of them apply the graph neural networks with high dimensional and sparse edge attributes into real recommendation systems, which process hundreds of millions of nodes and edges in seconds.

\subsubsection{Edge Attributes in GNN}
Besides node and topological information, edge attributes also play an important role in structure similarity learning. Existing methods use the edge attributes in two different ways based on subgraph and path respectively in the network. 
They aggregate the edges and nodes features by message passing.
generate the path representation between target nodes.
CensNet \cite{jiang2019censnet} uses graph convolution network with edge-node switching without the interaction of edges, which is hard to capture topological structure information through paths. GAS
 \cite{li2019spam} exploit edge information by concatenating the edges with their neighbor nodes, and the model proposed by \cite{gong2019exploiting} is able
to incorporate multi-dimensional positive-valued edge
features. It eliminates the limitation which can handle only binary edge indicators and one dimensional edge features. TALP~\cite{talp} predicts target links based on graph attention networks, which learn the type-aware vectors and type-fusion vectors associated with user nodes. However, the edge type in Heterogeneous graph is different from edge attributes, it's easy to observe the relationships between user behaviors from edge attributes pattern instead of edge vectors. Considering the above problem, we propose a novel attributed graph neural network to capture topological and sequential pattern for high-dimension sparse edge features.


\section{Attributed Graph Convolution Networks}
In this section, we introduce the new framework \textbf{A}ttributed \textbf{G}raph \textbf{C}onvolution \textbf{N}etworks (AGCN) including the symbolic definition of the problem, subgraph extraction method and detailed architecture.

\subsection{Problem Statement}
Given a graph $G=(V,E)$, $V$ represents the node set and $E$ is the edge set. $V_i\in \mathbb{R}^{|V| \times d}$ represents d-dimensional attribute vector of the $i^{th}$ nodes. $E_{ij} \in \mathbb{R}^n$, $i, j\in \{1, 2, ..., |V|\}$ is the n-dimensional attributed vector of the edge between node $i$ and node $j$, and $E_{ijk}$ denotes the edge $E_{ij}$ with k attributes, such as timestamp and category. Further more, we define $G_{(i,j)}^h$ noted as an enclosed subgraph with h-hop neighbors around the node $i$ and node $j$ and a path $P_{(i,j)}^h$ between node $i$ and $j$ with length h. The two nodes are treated as target nodes. Link prediction task is aimed to learn a model from the subgraphs and paths which can predict the probability of links between arbitrary two nodes.

\subsection{Subgraph Extraction} \label{sec:subgraphextract}
 In this work, we apply subgraph-based method in recommendation systems. Nodes in the neighborhood are stored in the same part to reduce data communication time. It makes industrial recommendation systems capture node representations from a large scale graph efficiently. This reduction not only saves the cost for the industry, but also updates the embeddings of users and products more quickly. 
Here we mainly refer to the subgraph extract methods used in \cite{zhang2017weisfeiler}. Given two nodes $i$ and $j$, extracting a h-hop subgraph $G_{(i,j)}^h$ by Algorithm~{\ref{alg:subgraph extraction}}.
\begin{algorithm}
	\caption{Subgraph Extraction}
	\label{alg:subgraph extraction}
	\begin{algorithmic}[1]
		\REQUIRE ~~\\ 
		{Graph $G$, target node pair $(i,j)$, hop number $h$, a node's maximal 1-hop neighbors $a$ } 
		\STATE $V_{(i,j)}=\{i,j\}$ 
		\STATE $temp=\{i,j\}$
		\STATE $b=0$ 
		\WHILE{$b < h$}
		\STATE for node $v$ in $temp$:\\ 
		    sample 1-hop neighbors $s(v)$, and satisfy $|s(v)|\leq a$
		\STATE $temp=(\cup s(v)) \verb|\| V_{(i,j)}$
		\STATE $V_{(i,j)}=V_{(i,j)}\cup temp$
		\STATE $b=b+1$
		\ENDWHILE
		\STATE according to $V_{(i,j)}$ and $G$, 
		generate edges set $E_{(i,j)}$\\
		generate nodes matrix $\bf{V_{(i,j)}}^{|V_{(i,j)}| \times m}$ \\
		generate edge tensor $\bf{E_{(i,j)}}^{|V_{(i,j)}| \times |V_{(i,j)}| \times n}$
		\RETURN $G_{(i,j)}^h=G(V_{(i,j)}, E_{(i,j)})$
	\end{algorithmic}
\end{algorithm}

\subsection{Model Architecture}
\label{sec:model}
\begin{figure}[t]
	\includegraphics[width=0.45\textwidth]{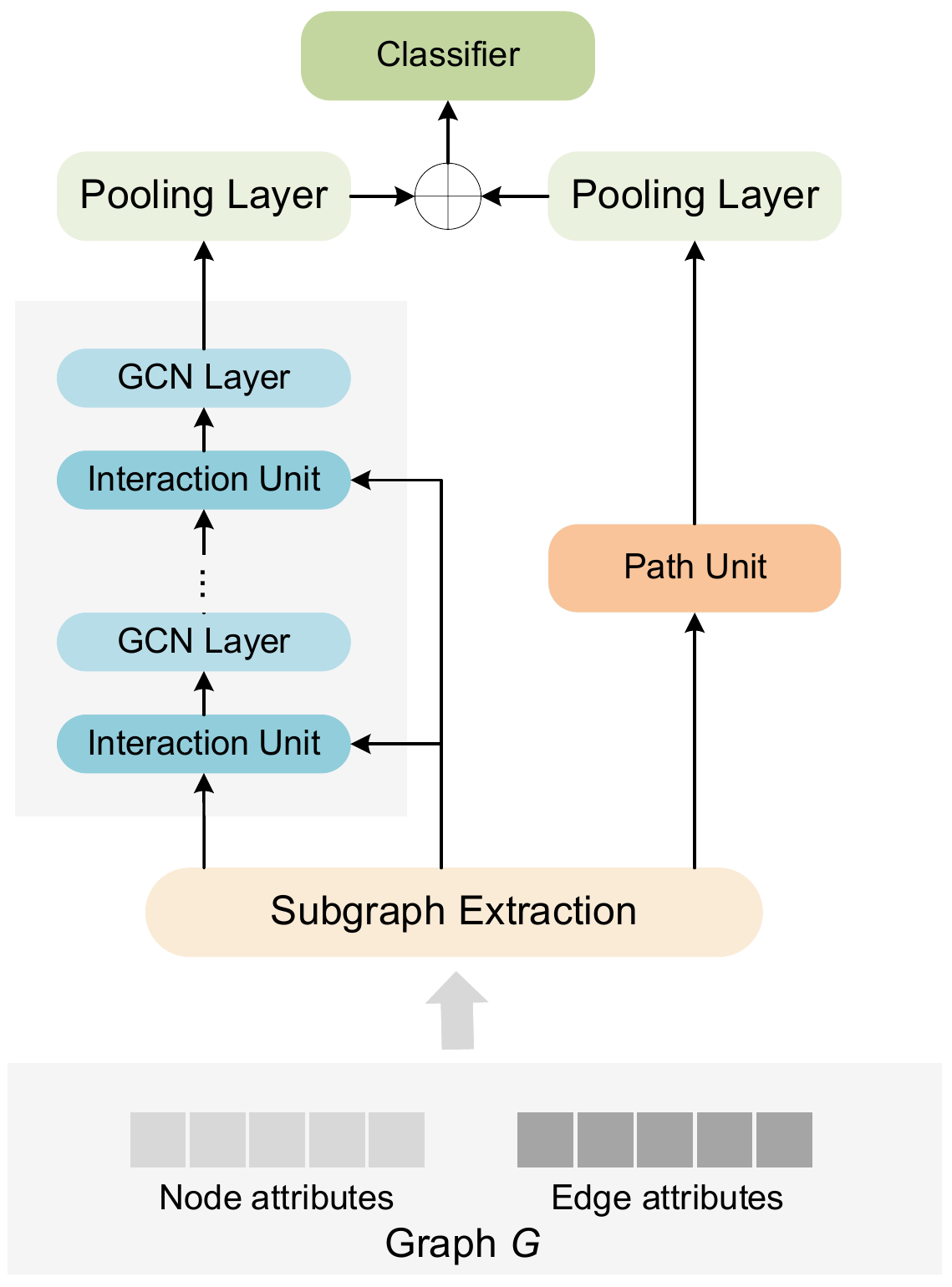}
	\caption{The architecture of Attributed graph convolution networks.}
	\label{fig:gcn} 
\end{figure}

The architecture of AGCN is shown in Figure~{\ref{fig:gcn}}. The network has two modules which respectively extract and process two kinds of structure information, subgraph and path. The main pipeline is described as the following. Firstly, We sample subgraphs around node $i$ and $j$ which satisfy the neighbor rules and obtain vertex and edge matrix. Each node and edge with attributes is encoded into the same feature space through two-level projection. This efficient two-level projection method is proposed here to reduce complexity caused by high-dimensional and sparse data in practice. Then we use the above interaction unit and gcn layers to add topological structures information. Node representations aggregate more and more information through the combination of interaction unit and GCN layer. Finally, we concatenate the node embeddings with the output of path unit to predict the link probability. The detailed descriptions are listed as follows.

\subsubsection{Interaction Unit} 
\begin{figure}[!ht]
	\includegraphics[width=0.45\textwidth]{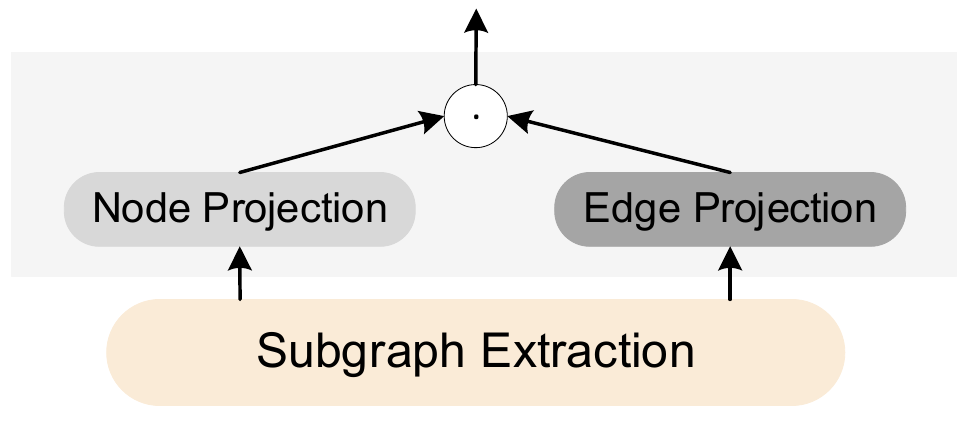}
	\caption{The detailed structure of interaction unit. The inputs are consist of node and edge representation. Two-level projections(node and edge) make raw features encoded into the same feature space.}
	\label{fig:IPU}
\end{figure}

Node-edge pairs are basic elements of topological structures in a subgraph, and the quality about the interaction of nodes and edges in pairs directly affects the final representing of the topological structure. As the attribute vectors are often very high-dimensional and sparse,  roughly crossing the attributes is prohibitive for it high cost. We propose a two-level projection method here ,as shown in Figure~{\ref{fig:IPU}}. In the first step, nodes attributes and edges attributes are projected to the same interaction space. And in the second step, element-wise product is adopted to complete the interaction. Assume the pair contains node $V_{x\cdot}$ and edge $E_{xy\cdot}$, the interaction process is as following

\vspace{-0.1cm}
\begin{equation}
	\label{eqn:vx'}
	\mathbf{V}_{x\cdot}^{l'}=\mathbf{E}_{xy\cdot}\mathbf{W}^{l}_{E} \bigodot \mathbf{V}_{x\cdot}^{l}\mathbf{W}^{l}_{V}\ .
\end{equation}

 $l$ means the $l^{th}$ layer of the graph neural network. $W_{E}^l$ is a trainable $n\times F^{l'}$ matrix, which projects $E_{xy}$ to a new $F^{l'}$-dimension latent space $\mathbb{R}^{F^{l'}}$. $V_{x}^l$ means the information of node $V_{x}$. After $l$ layers networks, $W_{V}^l$ has also the same function with $W_{E}^l$ and projects $V_{x\cdot}^l$ to the same latent space $\mathbb{R}^{F^{l'}}$. $\bigodot$ means element-wise production operation to combine two-level embeddings together.

\subsubsection{GCN Layer}
GCN layers aggregate the projected node-edge pairs in the neighborhood. The set of neighbors of node $y$ is denoted as $\mathcal{N}_{(y)}$. According to GCN ~\cite{GCN}, the update formulation of node $y$ is on the following.

\vspace{-0.1cm}
\begin{equation}
	\label{eqn:ag}
	\mathbf{V}_{y}^{l+1}=f(\frac{1}{|\mathcal{N}_{(y)}|+1}(\mathbf{V}_{y}^{l}\mathbf{W}^{l}_{V}+\Sigma_{x\in \mathcal{N}_{(y)}}\mathbf{V}_{x}^{l'})\mathbf{W}^l),
\end{equation}

where $W^l$ is a trainable parameters matrix, and $f$ is a point-wise nonlinear activation function.

Note that one GCN layer aggregates node-edge pairs in the neighborhood, and multiple GCN layers combine the h-hop pairs to represent the topological structures implicitly.


\subsubsection{Path Unit}
\begin{figure}
	\centering
	\includegraphics[width=0.4\textwidth]{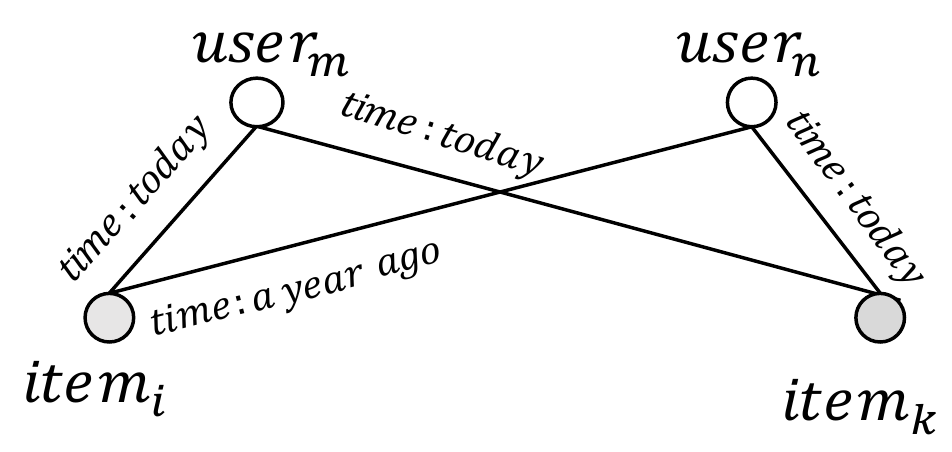}
	\caption{Examples of path attributes. In this figure, two paths are listed: one is $[item_i, time\colon today, user_m, time\colon today, item_k]$. The other path is $[item_i, time\colon a year ago, user_n, time\colon today, \\item_k]$.}\label{fig:route}
\end{figure}

This unit focuses on the presenting of paths that consist of edges. However in practice, high-dimensional and sparse attribute vectors are common and some dimensions often have insufficient samples, which may lead bad training. We propose a simple and effective way to deal with the problem. See Figure~{\ref{fig:route}}, assuming that the two users are same, whether the clicks happen in the same day can provide different information for link prediction of the two items. Intuitively $item_{i}$-$user_{m}$-$item_{k}$ contributes more to the link existence between the two items.

Denoting $r_{ij}^p$ as one of the paths between two target nodes $i$ and $j$ of a subgraph, which is as following

\vspace{-0.1cm}
\begin{equation}
	\label{eqn:rij}
	r_{ij}^p : [E_{i_{1}i_{2}\cdot}^p,E_{i_{2}i_{3}\cdot}^p,\cdots,E_{i_{a}j\cdot}^p],
\end{equation}

where $i_{0}=i$ and $a$ is the length of path $r_{ij}^p$. Further more, $E_{i_{h}i_{h+1}}^p$ can be divided into groups, where each group means a kind of edge attribute, such as time, trigger item, scenarios and so on. Each kind of attributes should be one hot format and $E_{i_{h}i_{h+1}}^p$ is as following

\vspace{-0.1cm}
\begin{equation}
	\label{eqn:Eij}
	E_{i_{h}i_{h+1}}^p=[f_{h1}^p, f_{h2}^p,\cdots,f_{hb}^p],
\end{equation}

where $b$ is the number of attribute groups and $f_{hl} \in \{0,1\}^{n_l}$ means attribute value, $n_l$ is the dimension of this kind of edge attribute. Path unit proposes a group-wise inner product method to encode path attributes, 

\vspace{-0.1cm}
\begin{equation}
	\label{eqn:rijeocode}
	r_{ij}^p = [\prod_{h=1}^af_{h1}^p,\prod_{h=1}^af_{h2}^p,\cdots,\prod_{h=1}^af_{hb}^p].
\end{equation}

After path unit, each path between two target nodes are encoded into a vector with value 1 or 0. In the pooling layer, sum pooling is adopted to further encode all the paths.

\subsubsection{Loss Function}
We formulate the link prediction problem as a subgraph classification problem. The objective function is as follows.
\vspace{-0.1cm}
\begin{equation}
	\begin{aligned}
		\label{eqn:loss}
		\min\ -\frac{1}{\# G_{(i,j)}^h}\sum\limits_{G_{(i,j)}^h\in G} (y(G_{(i,j)}^h)logM(G_{(i,j)}^h)\\
		+(1-y(G_{(i,j)}^h)log(1-M(G_{(i,j)}^h)))\ ,
	\end{aligned}
\end{equation}

where 
\begin{center}
	\begin{equation}
		\label{eqn:ygraph}
		y(G_{(i,j)}^h)=\left\{
		\begin{aligned}
			0  & , & E_{ij\cdot}=\bf{0} \\
			1 & , & \ otherwise 
		\end{aligned}
		\right.
	\end{equation}
\end{center}

\section{Experiments}\label{sec:exp}

\subsection{Datasets}
 we conduct experiments on two industrial datasets and one released dataset. These industry datasets provide richer edge attributes over public datasets and follow the real user interest distribution. The framework is applied in real recommendation system called Xianyu which is the largest second-hand transaction platform in China. Due to missing edge attributes in most public graph-structured datasets, we use Epinions Dataset \footnote{http://www.trustlet.org/downloaded\_epinions.html} to build the attributed graph.



\subsubsection{XianyuRaw}
This dataset is derived from Xianyu platform with
 2,399,900 users and 1,846,099 items and 20,283,822 click edges between them. Clicks often happen under different conditions, such as time difference, scenarios(feeds, search etc.), trigger items and queries. Time difference means the gap between today and the time behaviors happened. These attributes play an important role on user interest mining. We define user-click-item bipartite graph to predict the possibility of interactions. 

\subsubsection{XianyuSparse}
 XianyuSparse contains 92,853 users and 91,848 items and 171,124 click edges. XianyuSparse aims to construct a graph where paths between nodes are very sparse. All the samples from XianyuSparse have no more than 3 paths. The detailed statistical results are shown in  Table~{\ref{tab:xyroutes}}.
\begin{table}[!htb]
	\centering
	\caption{Average path samples in Xianyu.}
	\label{tab:xyroutes}
	\begin{tabular}{ccc}
		\toprule
		Testing Set & XianyuRaw & XianyuSparse \\
		\midrule
		positive samples & 16.71 & 1.99 \\
		negative samples & 20.43 & 1.62 \\
		\bottomrule
	\end{tabular}
\end{table}

\subsubsection{Epinions}
this dataset contains 49,290 users who rated a total of 139,738 different items. The users and items can be seen as nodes in the graph. The number of rated user-item pairs is 664,824, which is a type of edge of the graph. The rating values are integers ranging from 1 to 5 who can be seen as the attributes of the edges. Another type of edges are 487,181 issued trust statements about users. We define $U2U$ (i.e. recommending an user to another user, which is a classic problem in recommendation systems) as a link prediction problem in Epinions bipartite graph. Training and testing samples are extracted from this graph, and the average path num of positive samples is 2.8 and that of negative samples is 1.3.


\subsubsection{Baselines}
This section introduce some classic methods as the baselines, including heuristic method(Jaccard, Adamic-Adar) and supervised learning algorithm(SEAL). 
Besides the above methods, we utilize edge attributes by concatenation as a baseline.
\begin{itemize}
    \item \textbf{Jaccard}: $|\mathcal{N}(x) \cap \mathcal{N}(y)|$. Denote $\mathcal{N}(x)$ as the set of 1-hop neighbors of node $x$. Jaccard assumes a positive relationship between link and 1-hop neighbors. This first-order method is shown as follows.
    \item \textbf{Adamic-Adar}:$\Sigma_{z \in |\mathcal{N}(x) \cap \mathcal{N}(y)|} \frac{1}{log(|\mathcal{N}(z)|)}$ This is a second-order method. It is widely used in industry due to its efficiency and effectiveness. The formula is listed as below.
    \item \textbf{SEAL}: SEAL is the first work which applies GNN to link prediction task \cite{zhang2018link}. SEAL exploits graph structure information and nodes information, and it outperforms heuristic methods and latent feature methods.
    \item \textbf{GAS}: GAS\cite{li2019spam} is proposed for link classification. We find that the proposed method about edge attributes can be applied in industry. They concatenate the edge attributes with neighbors to represent the topological structures. 
\end{itemize}

%


\subsection{Experimental Settings}

\subsection{Offline Evaluation}
Table~{\ref{tab:offline}} lists the performance of the proposed model and other methods, and AGCN obtains the best performance. Detailed results and analysis are listed as follows.
\begin{table}[!htb]
	\centering
	\caption{Offline AUC metrics in three datasets.}
	\label{tab:offline}
	\begin{tabular}{c|ccc}
		\toprule
		Method & XianyuRaw & XianyuSparse & Epinions\\
		\midrule
		Jaccard & 0.5161 & 0.3242 & -\\
		Swing & 0.3645 & 0.4743 & 0.6798 \\
		SEAL& 0.6536 & 0.5000 & 0.6843 \\
		Concat & 0.6544 & 0.6867 & - \\
		AGCN & \textbf{0.6775} & \textbf{0.7100} & \textbf{0.7035} \\
		\bottomrule
	\end{tabular}
\end{table}

\subsubsection{Performance Comparison}

As mentioned above, the baseline methods can be divided into two categories, heuristic methods (Jaccard and Adamic-Adar) and supervised methods (SEAL and Concat). For heuristic methods, their link prediction function is often based on human assumptions, and the parameters of the model are manually defined by humans. On the dataset XianyuRaw and XianyuSparse , given the complexity and sparseness of the data, the heuristic methods include too many human defined factors, so the Jaccard and Adamic-Adar methods perform the worst. For supervised methods, the parameters of the model are learnable, so compared to heuristic methods, supervised methods have stronger learning capabilities and perform better. The supervised methods SEAL and Concat here are both GNN-based methods. When they model the information on the graph, they only exploit structure information and ignore the attributes of the edges in the graph. When the edges on the graph are very sparse, the structure of the graph becomes similar and difficult to distinguish, so the performance of the model will decrease. Different from the above two methods, our proposed AGCN method can consider the interaction of edges and nodes at the same time, so as to better model the topological structure on the graph. As shown in Table~\ref{tab:offline}, our AGCN can achieve far better results than the baseline methods on all three datasets.

In summary, most industrial datasets can be represented in the form of graphs, but due to their distribution and sparseness, it is difficult for humans to manually define appropriate rules and assumptions. Therefore, traditional heuristic methods cannot achieve good results in these dataset. Our proposed AGCN method can learn the various rules from the respective datasets, and exploit the information of the node-edge pair and path.  Therefore, it can play a greater role in the link prediction tasks based the above datasets.


%
\begin{figure}[!htb]
	\includegraphics[width=0.45\textwidth]{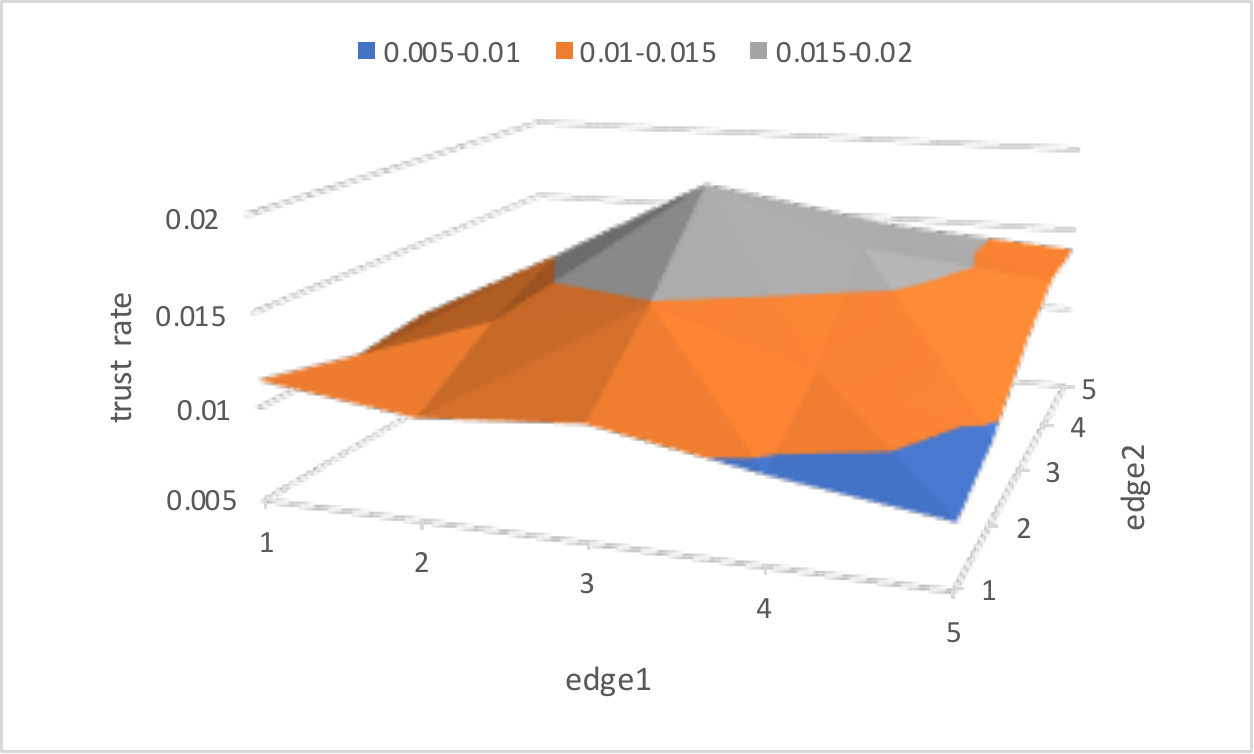}
	\caption{A 3-dimensional surface graph which represents the relationship between edges and trust rate in the Epinions dataset. Edge1 and edge2 means two kinds of edges in one path, like $[user1,item,user2]$. }\label{fig:epedge}
\end{figure}

\subsection{Impact of Edge Attributes}

\begin{figure*}
    \centering
    \includegraphics[width=\linewidth]{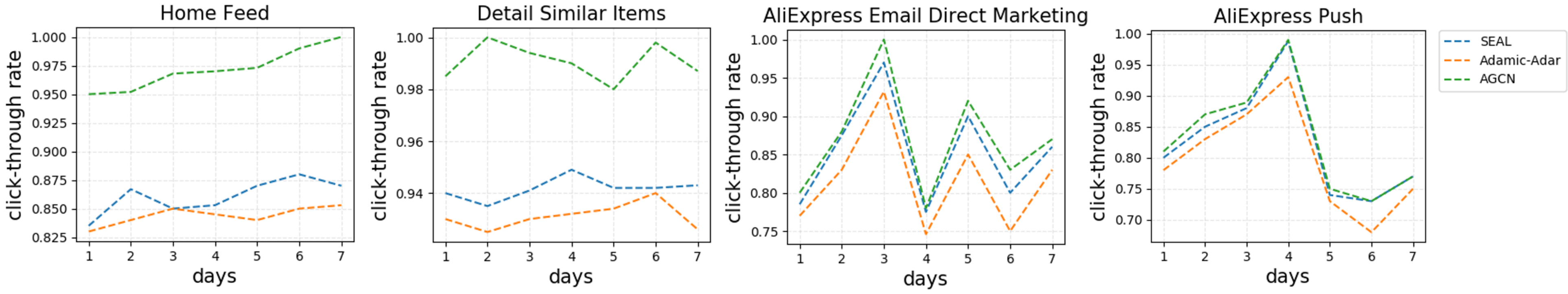}
    \caption{Online performance of Adamic-Adar, SEAL and AGCN in four scenarios within one week: home feed and detail similar items in Xianyu, email direct marketing and push in AliExpress.}
    \label{fig:my_label}
\end{figure*}


 We extract all the paths which length is equal to 2. In the Epinions bipartite graph, the path is set as user1-item-user2. Each path can be divided into two edge: edge1(user1-item) and edge2(user2-item), and the label of user1-user2 is assigned to this path. 
 Figure~{\ref{fig:epedge}} demonstrates that edge attributes are related with the links between users. The trust rate reflect the approximate capacity of models. It's clear that this relationship needs to learn through nonlinear models and different edge combinations({\it i.e.} paths) have different contributions to the prediction. However, Heuristic methods like common neighbors and GNN-based methods can not explicitly use path feature. In our work, we not only learn node representation, but add the path information into node-level and path-level representation.

\subsection{Online Experiments}

\subsubsection{A/B Testing}
The proposed AGCN was deployed in two industry APPs: XianYu which is the largest second-hand transaction platform in China, and AliExpress which is an international e-commerce platform. In XianYu, two scenarios adopted AGCN method: Feeds in homepage and Similar Items in item detail page. Feeds is that users can browse endless items if they want, Similar Items is a scenario that some items will be recommended if the user clicks an item. In AliExpress, AGCN is deployed in Email Directing Marketing scenario and Push scenario. Email Directing Marketing and Push will send some items or messages to the user through email or phone message channel trying to attract them into AliExpress.

In the four scenarios, we choose two kinds of nodes, user and item. The click action is the type of edge with many attributes. We predict the link of user-item(U2I) in Feeds and item-item(I2I) in Detail Similar Items. The difference between I2I and U2I is that I2I means find similar items for a target item and it is treated as the link prediction problem. I2I is an effective method in industrial online matching stage. U2I aims to capture users' interest. Training data is collected from historic data of Xianyu or AliExpress. The label of user-item pairs is set as 1 when the user clicked the items while the label of item pairs is set as 1 when similar items are recommended and clicked after the trigger item clicked. We evaluate the real-time performance through online A/B testing platform.

\begin{table}[!htb]
	\centering
	\caption{Online A/B testing performance of feeds and detail similar items. PU is path unit and IU means interaction unit. Online metrics is a comprehensive measure to evaluate the global online performance including pctr, pcvr, ipv etc.}
	\label{tab:onlineauc}
	\begin{tabular}{c|c|ccc}
		\toprule
		Scenario & Method & Online  Metrics\\
		\midrule
		&Baseline & +0.0\% \\
		Feeds & $\text{AGCN}_\text{PU}$ & +10\% \\
		&$\text{AGCN}_\text{PU\&IU}$& +15\% \\
		\midrule
		&Baseline & +0.0\% \\
		Detail Similar Items&$\text{AGCN}_\text{PU}$ & +6\% \\
		&$\text{AGCN}_\text{PU\&IU}$ &  +10\%  \\
		\bottomrule
	\end{tabular}
\end{table}

%


We deploy AGCN in Xianyu and AliExpress, and the daily online metrics are shown in Figure~{\ref{fig:my_label}}. The figure leads us to the conclusion that in four scenarios, AGCN achieves the best performance stably within one week and shows the superiority in the industrial recommendation systems. The specific improvement shows in Table ~\ref{tab:onlineauc}. It's clear from the table that AGCN with path unit outperforms 6\% $\sim$ 10\% online metrics over baseline method. AGCN with path unit and interaction unit also improve about 10\% $\sim$ 15\%.
Therefore, it can be concluded that AGCN have more powerful modeling capabilities for high-dimensional sparse edge attributes.


\begin{figure}[!hbt]
	\includegraphics[width=\textwidth]{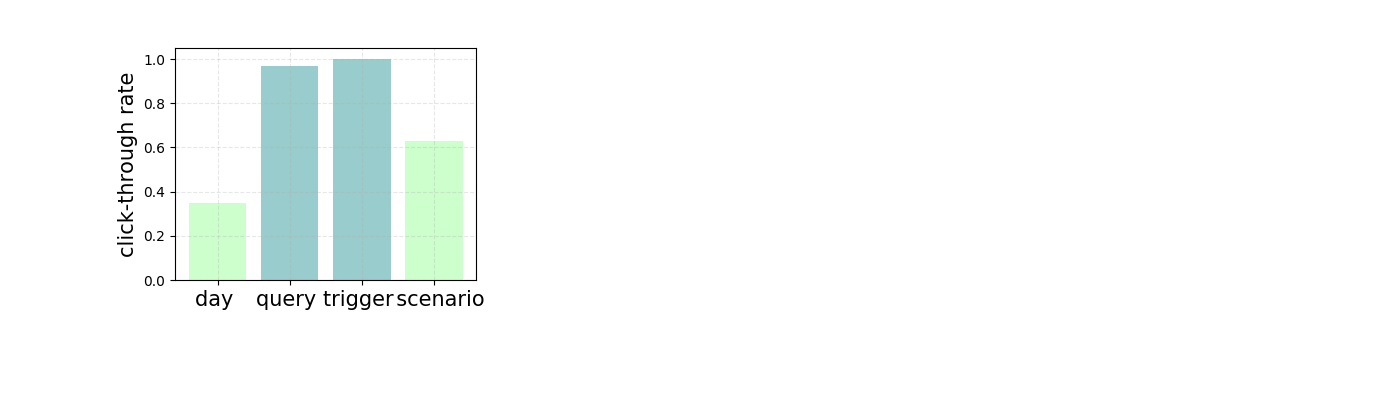}
	\vspace{-1.5cm}
	\caption{Online click-through rate with the same attributes of paths in the sparse graph data. In this graph, 70\% path lengths are equal to 1. The same trigger in path lead to the largest click-through rate.}\label{fig:sparse2}
\end{figure}
\subsubsection{Sparsity Analysis}
We further analyze the reason of the improvement. We collect the online log of the four scenarios, and we find it suffers very high sparsity. The proportion of the two target item pairs whose path number equals to one is as high as 70\% of the total. It means that most item pairs are almost the same from the view of graph structure, and they cannot be distinguished well only through heuristic methods. Further more, similar with Figure~\ref{fig:epedge}, we analyze the edge's impact on the pairs with only one path, and {\it Path unit} method in AGCN is adopted here to make the impact be visualized. As shown in Figure~\ref{fig:sparse2} (The results have been normalized), the abscissa axis means the edges in the path happen in the same day, or in same trigger, or in same query, or in same scenario, and the vertical axis means click-through rate. As shown in Figure~\ref{fig:sparse2}, we can see that edge attributes have different impacts on link prediction(click-through rate), and this demonstrates the effectiveness of {\it Path Unit} and is one of the key reasons why AGCN outperforms the baseline methods.

\section{Conclusions}
In this paper, we propose an attributed Graph Convolution Network to incorporate edge attributes when apply graph neural network in the large-scale sparse graph-structured datasets. Our proposed framework includes three parts. Firstly, we propose a two-level projection to decompose the topological structures to node-edge pairs and project them into the same interaction feature space. Then, we apply the multi-layer GCN to combine the projected node-edge pairs to capture the topological structures. Finally, we formulate the link prediction task as a subgraph classification problem. A large number of offline and online experiments prove that our proposed AGCN is far better than the baseline methods.


\newpage
\bibliographystyle{aaai.bst}
\bibliography{agcn.bib}

\newpage
\appendix

%
%

\end{document}